# Subsurface hydrogen storage controlled by small-scale rock heterogeneities

Zaid Jangda[1,*], Hannah Menke[1], Andreas Busch[2], Sebastian Geiger[3], Tom Bultreys[4], Kamaljit Singh[1]

[1]Institute of GeoEnergy Engineering, Heriot-Watt University, EH14 4AS, Edinburgh, United Kingdom

[2]Lyell Centre, Heriot-Watt University, EH14 4AS, Edinburgh, United Kingdom.

[3]Department of Geoscience and Engineering, Delft University of Technology, 2628 CN Delft, Netherlands

[4]UGCT/PProGRess, Department of Geology, Ghent University, 9000 Ghent, Belgium

*Corresponding author

Email addresses: zj21@hw.ac.uk (Z. Jangda), h.menke@hw.ac.uk (H. Menke), a.busch@hw.ac.uk (A. Busch), s.geiger@tudelft.nl (S. Geiger), tom.bultreys@ugent.be (T. Bultreys), k.singh@hw.ac.uk (K. Singh)

## Abstract

Subsurface porous rocks have the potential to store large volumes of hydrogen ($H_2$) required for transitioning towards a $H_2$-based energy future. Understanding the flow and trapping behavior of $H_2$ in subsurface storage systems, which is influenced by pore-scale heterogeneities inherent to subsurface rocks, is crucial to reliably evaluate the storage efficiency of a geological formation. In this work, we performed 3D X-ray imaging and flow experiments to investigate the impact of pore-scale heterogeneity on $H_2$ distribution after its cyclic injection (drainage) and withdrawal (imbibition) from a layered rock sample, characterized by varying pore and throat sizes. Our findings reveal that even subtle variations in rock structure and properties significantly influence $H_2$ displacement and storage efficiency. During drainage, $H_2$ follows a path consisting of large pores and throats, bypassing the majority of the low permeability rock layer consisting of smaller pores and throats. This bypassing substantially reduces the $H_2$ storage capacity. Moreover, due to the varying pore and throat sizes in the layered sample, depending on the experimental flow strategy, we observe a higher $H_2$ saturation after imbibition compared to drainage, which is counterintuitive and opposite to that observed in homogeneous rocks. These findings emphasize that small-scale rock heterogeneity, which is often unaccounted for in reservoir-scale models, can play a vital role in the displacement and trapping of $H_2$ in subsurface porous media.

## Introduction

With the increasing global demand for clean and sustainable energy sources, $H_2$ has emerged as a promising solution as a low-emission energy-carrier,[1–3] particularly for hard to electrify decarbonizing sectors like heavy industry, long-haul transportation, and seasonal storage.[4] The surge in the worldwide interest in $H_2$ is evident from the announcement of more than 1000 large-scale $H_2$ production, supply, and infrastructure projects globally, with over 75% of them targeting full or partial commissioning by 2030.[5] These projects represent a total of $320 billion in announced investments.[5] It is predicted that if clean $H_2$ is scaled up globally, the $H_2$ industry has the potential to generate $2.5 trillion in annual revenues and 30 million jobs, along with 20% global emissions reductions by 2050.[6] Therefore, the demand for $H_2$ is expected to be substantial, with estimates reaching 2500 TWh for Europe by 2050.[7] To accommodate such a demand, efficient $H_2$ storage solutions are essential, with projected storage requirements of 250-1000 TWh in Europe by 2050[7] and potentially several gigatons globally.[8]

Underground storage in depleted hydrocarbon reservoirs or aquifers emerges as a potential solution that provides the required storage capacity for these immense volumes of $H_2$.[3,9–12] However, for the successful implementation of underground hydrogen storage (UHS) in porous reservoirs, it is crucial to understand the displacement of $H_2$ during injection and withdrawal cycles.[13–17] To develop this understanding, it is important to analyze and quantify the interactions between $H_2$ and the subsurface rocks and fluids and investigate the intricate pore-scale mechanisms that govern $H_2$ displacement within these subsurface rocks.[8,18–20]

Experimental studies conducted to unravel these pore-scale phenomena have provided initial insights into the flow, transport, and trapping of $H_2$ in sandstone rocks. Wettability studies have generally concluded that water-wet conditions would prevail in an $H_2$-brine system.[19–25] Recently, pore-scale fluid imaging experiments have been conducted to visualize and quantify the $H_2$ saturations and different pore-scale mechanisms occurring within rocks during and after $H_2$ injection and withdrawal. Results from these studies show that there will be a loss of $H_2$ in the subsurface due to capillary trapping,[14,20,25,26] potential dissolution of $H_2$ in the imbibing brine,[20] and that re-arrangement of $H_2$ could occur in the pore-space due to Ostwald ripening.[8] These studies offer valuable insights into the pore-scale phenomena that would arise during the implementation of large-scale UHS. However, it is important to note that almost all these investigations were conducted on homogeneous rocks and (except[20]) were not carried out under representative subsurface temperature and pressure conditions.

Most subsurface reservoirs are heterogeneous, meaning that pore and throat sizes, porosity, and permeability vary over a wide range of magnitudes and length scales.[27–31] In heterogeneous rocks, capillary pressure effects arise at boundaries between regions with distinctly different properties, i.e., with different permeability or porosity.[29] Heterogeneity in a geological formation can promote channeling of fluid flow along preferential flow paths that have high permeability, which leads to the bypassing of potentially large reservoir volumes and more complex displacement patterns.[27,29,30,32,33] Even small variations in permeability can alter fluid displacement patterns considerably.[29,34] Results from core-scale experiments showed that in a heterogeneous rock, the formation of high water saturation channels during imbibition could result in significant $H_2$ trapping during $H_2$ withdrawal, as a portion of the $H_2$ phase may be bypassed.[33] Results from a recent two-dimensional micromodel study,[35] show that the change of flow direction from coarse to fine and fine to coarse sections of the micromodel significantly influenced the trend of the averaged capillary pressure curves and the remaining fluid saturation. Permeability heterogeneity is also expected to play a role in $H_2$ entrapment and recovery efficiency at the reservoir scale, with a recent simulation study showing 7% incremental $H_2$ recovery from a homogeneous reservoir compared to a heterogeneous reservoir.[36] Pore-scale heterogeneity is hence expected to play a critical role in both, the storage capacity of $H_2$ and the volume of $H_2$ that could be produced back from subsurface storage systems.

In this study, we conducted $H_2$ flow experiments using a heterogeneous sandstone rock to investigate the impact of pore-scale heterogeneity on $H_2$ displacement and trapping. The experiments were carried out at temperature and pressure conditions of 50°C and 10 MPa respectively, using a custom-designed flow apparatus, with the rock imaged *in situ* in an X-ray micro-computed tomography (μCT) scanner. We conducted two separate experiments to replicate two different scenarios that could arise during $H_2$ injection (a drainage process) and withdrawal (an imbibition process) from rock samples. Each experiment consisted of two cycles of drainage and imbibition. During drainage, $H_2$ was injected into the initially brine saturated rock sample from the top to simulate $H_2$ injection into a subsurface reservoir, while during imbibition, brine was injected from the bottom to simulate the displacement of stored $H_2$ during its withdrawal from a subsurface reservoir. After each fluid displacement step, we visualize and quantify the $H_2$ saturation in the rock and highlight the effect of pore-scale heterogeneity on fluid displacement.

The magnitude of the pore-scale heterogeneity present in the rock used for our experiments is generally not considered when developing reservoir-scale models, where permeability variations are usually incorporated when the difference is larger than an order of magnitude.[31] However, our findings emphasize that $H_2$ flow and trapping is highly influenced by subtle variations in the rock structure and properties, even for permeability contrasts lower than an order of magnitude. Furthermore, we highlight the difference in the fluid distributions depending on the fluid injection strategy for each experiment performed in this work.

## Materials and methods

### *Equipment and materials*

A cylindrical Clashach sandstone rock sample with a diameter of 6 mm and length of 12.4 mm was used as the porous medium for our experiments. Clashach is a quarried sandstone from Scotland, primarily consisting of approximately 90% quartz and 10% K-feldspar. Its permeability ranges from $2 \times 10^{-13}$ to just over $1 \times 10^{-12}$ m$^2$ and its porosity from 12 to 18%.[37,38] These values are similar to the properties of sandstones found in depleted North Sea hydrocarbon reservoirs[39]. Experimentally measured water permeability for the rock sample used for our experiments was found to be $1.1 \times 10^{-12}$ m$^2$. A brine solution (de-ionized water doped with 4 wt.% potassium iodide to provide effective X-ray contrast) was used as the aqueous (wetting) phase and high purity (>99.99%) $H_2$ (supplied by BOC) was used as the gas (non-wetting) phase. The brine solution used for the saturation of the rock and imbibition was pre-equilibrated with $H_2$ at the experimental conditions in a Hastelloy reactor (Parr Instruments Company), to mitigate the loss of $H_2$ in the rock sample due to the potential dissolution of $H_2$ in the brine.

Before starting the experiments, the rock sample was cleaned by immersing it in methanol under a fume hood for 20 hours and then dried in a vacuum oven at 100°C for 24 hours. The rock sample was wrapped in Teflon tape and aluminum foil and placed inside a Viton sleeve. The Viton sleeve containing the rock sample was then wrapped with a $H_2$ leak detection tape (Nitto), before fitting inside a custom-designed rock core-holder (rs systems). The core-holder was vertically fixed on a rotation stage inside an X-ray μCT scanner (EasyTom 150, RX Solutions) and connected to the flow system consisting of four syringe pumps (ISCO, three model 500D, and one model 100DX) and the Hastelloy reactor. Details of the rock sample preparation procedure and the experimental apparatus including the flow system can be found in our previous work.[20] A differential pressure transducer (Keller, PD-33X) was added to the experimental setup to record the pressure difference between the top and bottom of the rock sample during all the fluid displacement steps. An additional flow line was also added to the bottom of the rock sample for one of the two experiments. These modifications are shown in the flow diagram in Figure S1 (Supplementary data).

*Experimental procedure*

The central vertical section of the rock sample was initially scanned using X-ray µCT before injecting any fluids. This dry scan served as a reference for the subsequent wet scans (containing $H_2$ and brine) taken after each fluid displacement step. The scanning parameters for all the scans are provided in Table S1 (Supplementary data).

For each of the two experiments performed in this work, $CO_2$ gas was first flushed through the rock sample at a pressure of 0.2 MPa to displace any air present in the rock sample. The $CO_2$ flush was followed by injection of 100 pore volumes (PV) of the brine solution at a flow rate of 0.5 mL.min$^{-1}$ which was increased to 1 mL.min$^{-1}$, ensuring 100% brine saturation in the rock sample. The pore pressure inside the rock sample and the confining pressure around the Viton sleeve were then gradually increased in steps of 0.2 MPa to the experimental pressure of 10 MPa and 12 MPa respectively. The rock sample was then heated to 50°C and allowed to stabilize for one hour. Next, the brine inside the rock sample was completely displaced with 70 PV of $H_2$-equilibrated brine from the reactor at a flow rate of 0.5 mL.min$^{-1}$ to achieve full saturation of the rock sample with $H_2$-equilibrated brine. We then started $H_2$ injection (drainage) from the top of the rock sample at a flow rate of 0.05 mL.min$^{-1}$, corresponding to a capillary number (Ca) of $4.2 \times 10^{-9}$. Here, $Ca = \mu v/\gamma$, where $\mu$ is the viscosity of $H_2$ ($9.64 \times 10^{-6}\ Pa.s$)[40], $v$ is the velocity of the injected $H_2$ ($2.95 \times 10^{-5}\ m.s^{-1}$), and $\gamma$ is the interfacial tension between $H_2$ and water ($0.0683\ N.m^{-1}$)[41] at the experimental conditions of 50°C and 10 MPa.

$H_2$ injection was stopped after 15 PV and the rock sample was scanned. Following this, 5 PV of brine was injected (imbibition) from the bottom of the rock sample to displace the $H_2$ at the same flow rate of 0.05 mL.min$^{-1}$, corresponding to a Ca of $2.3 \times 10^{-7}$. Another scan was acquired after this displacement step. This drainage-scan-imbibition-scan cycle was then repeated. Figure S2 and Figure S3 (Supplementary data) show the schematic of the experimental steps for each of the two experimental strategies used in this work.

## Experimental strategies

Two separate experimental strategies were used. In the first strategy, the rock was connected to two flowlines at the bottom of the rock sample (Figure 1A and Figure 1B).

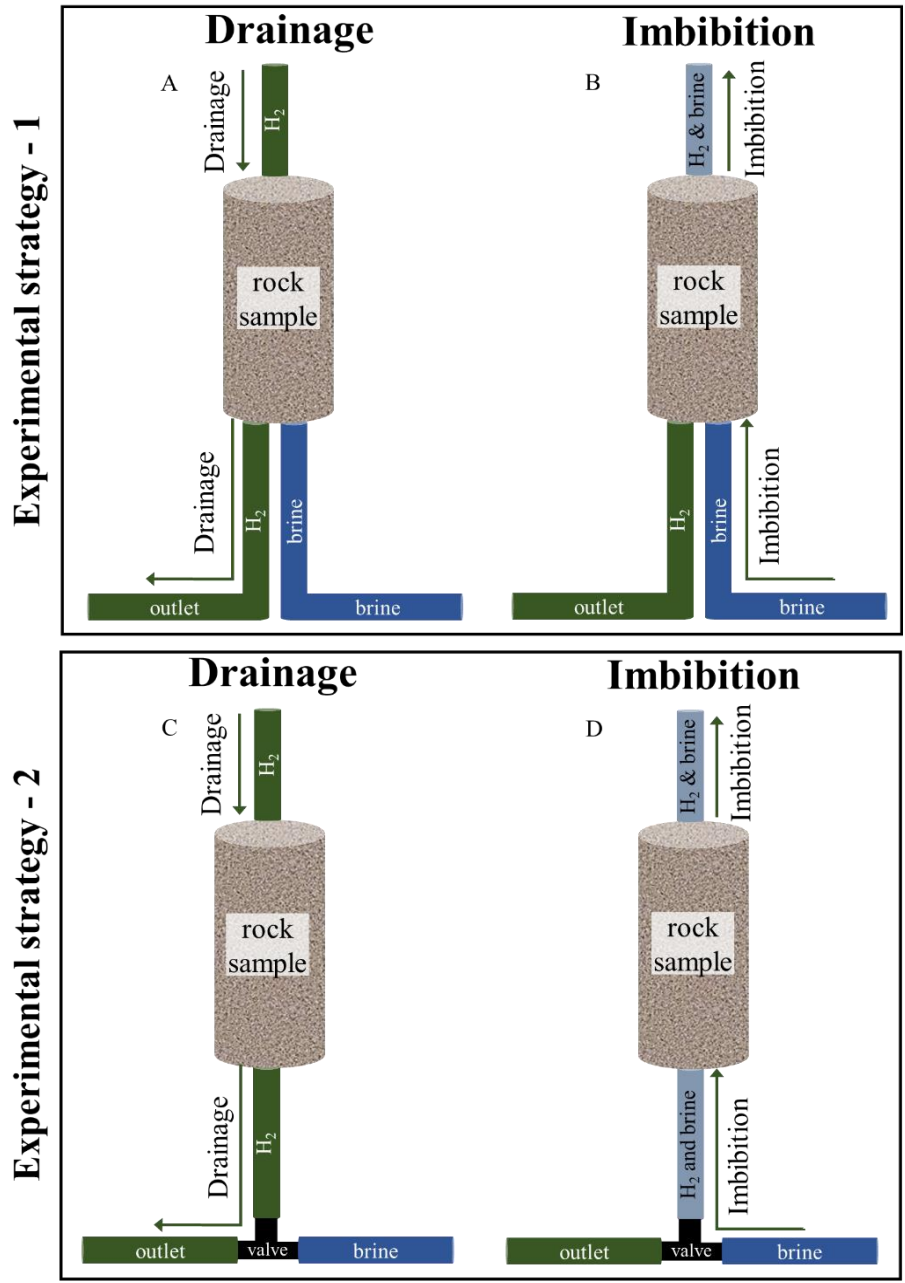

**Figure 1.** A simplified visualization of the flow lines used for the two experiments. (A) & (B) show Experiment 1 in which two separate flowlines at the bottom of the rock sample were used, one for $H_2$ to flow out from the bottom of the rock sample during drainage (A) and another for brine injection into the rock sample during imbibition (B). (C) & (D) show Experiment 2 in which a single line was used for $H_2$ to flow out from the bottom of the rock sample during drainage (C) and for brine injection into the rock sample during imbibition (D).

When the rock sample was initially saturated with $H_2$-equilibrated brine, both flow lines were filled with this brine. During drainage, when $H_2$ was injected into the rock sample from the top, one flowline at the bottom allowed the flow of brine and $H_2$ to the receiving pump, while the second flowline remained filled with brine (Figure 1A). This second flowline was then used to inject brine from the bottom during imbibition (Figure 1B). This setup ensured that there was no dead volume of $H_2$ in the flow line which would get injected into the rock sample during imbibition.

In the second strategy, only one flowline was connected to the bottom of the rock sample, serving as the outlet during drainage (Figure 1C) and the inlet during imbibition (Figure 1D). Consequently, the $H_2$ that accumulated in the bottom flowline during drainage was reintroduced into the rock sample at the commencement of brine injection. The slight variation in the experimental technique, which is generally used for experiments conducted on homogeneous rock samples,[20] significantly influenced the fluid distributions, due to the inherent heterogeneity present in the rock sample, as discussed in the Results and discussion section.

*Image processing and pore network extraction*

The raw data from each scan was reconstructed using the EasyTom Xact software after which the images were processed using Avizo (ThermoFisher scientific) software. To visualize a longer length of the rock sample, the dry rock sample was scanned (at a voxel size of 5 µm) at two different heights. These two images were then stitched together resulting in the visualization of an 11.6 mm long central vertical section of the rock sample. From this stitched image, a sub-volume corresponding to a length of 8.8 mm was selected to be scanned (at a voxel size of 7 µm) for all the wet scans. The voxel size was increased for the wet scans to ensure that we could image the fluid displacement in layers of varying pore and throat sizes within the rock sample in a single scan, while maintaining a good image resolution.

The image from the dry scan was resampled to match the length and voxel grid of the wet scans. Thereafter, all the images from the wet scans were registered to the image from the dry scan. A sub-volume corresponding to a length of 8.44 mm was selected for qualitative and quantitative analysis. All the images were then filtered using a non-local means filter[42] to remove noise. Image segmentation was then performed on all the images using a watershed algorithm[43] based on the grayscale intensity values. The image from the dry rock sample was segmented into rock grains and pores, while the $H_2$ phase was segmented from all wet scans. The segmented pores from the dry image served as a mask to obtain the brine phase for each of the $H_2$-segmented wet images. The image segmentation procedure is explained in detail in our previous work[20] and the threshold grayscale intensity values for each segmented image are provided in Table S2 (Supplementary data).

The segmented pore space from the dry scan was also used to extract a simplified pore network of the rock sample, consisting of spherical pores and throats. This pore network extraction was achieved through the application of the maximal ball algorithm,[44–46] which allowed us to get the pore and throat size distributions and information about the pores and throats occupied by $H_2$ after each fluid displacement step (i.e., drainage and imbibition).

*Sectional analysis*

In addition to analyzing the full sub-volume, we conducted a more detailed analysis of the rock sample by dividing it into three layers: bottom, middle, and top (Figure 2A). This division was based on the average pore areas of the 2D vertical slices. To calculate these averages, we used the 2D area of each pore in every vertical slice. A central section, hereafter referred to as the middle layer, was selected by identifying consecutive slices with an average 2D pore area of less than 0.01 mm² (2D equivalent pore radius ~ 84 µm). This division allows for a comprehensive examination of the interplay of fluid flow and trapping behavior within each layer, owing to the variations in pore and throat sizes (and therefore porosity and permeability).

Table 1 shows the lengths, the average 3D pore and throat radii, and the permeabilities (obtained numerically using GeoChemFoam[47,48]) of each layer.

**Table 1.** Lengths, pore, and throat radius (mean ± standard deviation) and permeability of the different layers of the core sample. Permeability values were obtained using the simpleFoam solver in GeoChemFoam 5.0

|  | **Full sample** 0-8.44 mm 1207 slices | **Bottom layer** 0-3.67 mm 524 slices | **Middle layer** 3.67-5.11 mm 206 slices | **Top layer** 5.11-8.44 mm 477 slices |
|---|---|---|---|---|
| **Avg. pore radius (µm)** | 31 ± 12 | 35 ± 14 | 28 ± 10 | 30 ± 11 |
| **Avg. throat radius (µm)** | 15 ± 7 | 17 ± 8 | 14 ± 6 | 15 ± 7 |
| **Permeability (m²)** | $1.0 \times 10^{-12}$ | $3.4 \times 10^{-12}$ | $5.1 \times 10^{-13}$ | $1.0 \times 10^{-12}$ |

These layers in the rock sample exhibit minor differences (<10 µm) in pore and throat radii. Even though the differences in the average pore and throat radii between the three layers are relatively small, they do exhibit permeability variations. Specifically, the bottom layer has a significantly higher permeability ($3.4 \times 10^{-12} \ m^2$) compared to the middle ($5.1 \times 10^{-13} \ m^2$) and top ($1.0 \times 10^{-12} \ m^2$) layers. Permeability variations of this magnitude and length-scales are challenging to upscale into large-scale reservoir models, and rocks containing small-scale heterogeneity are often grouped as a single hydraulic unit.[31] However, we observe that these small-scale heterogeneities contribute significantly to the $H_2$ movement and trapping in the rock sample.

Analysis of the 2D average pore area (represented by 2D equivalent pore radius in Figure 2B) for each slice shows the presence of a layer in the middle of the rock sample characterized by smaller pores (Figure 2B). When considering the 3D positions of all the pores and throats, we observe that the bottom layer consists of a significantly larger number of larger pores and throats (Figure S4 in Supplementary data). While the middle layer consists mostly of relatively smaller pores and throats compared to the bottom and top layers, there are a few large pores and throats present in the middle layer. These large pores and throats in the middle layer are likely to provide an interconnected pathway, facilitating the initial flow of $H_2$ between the top and bottom layers as discussed in the Results and discussion section.

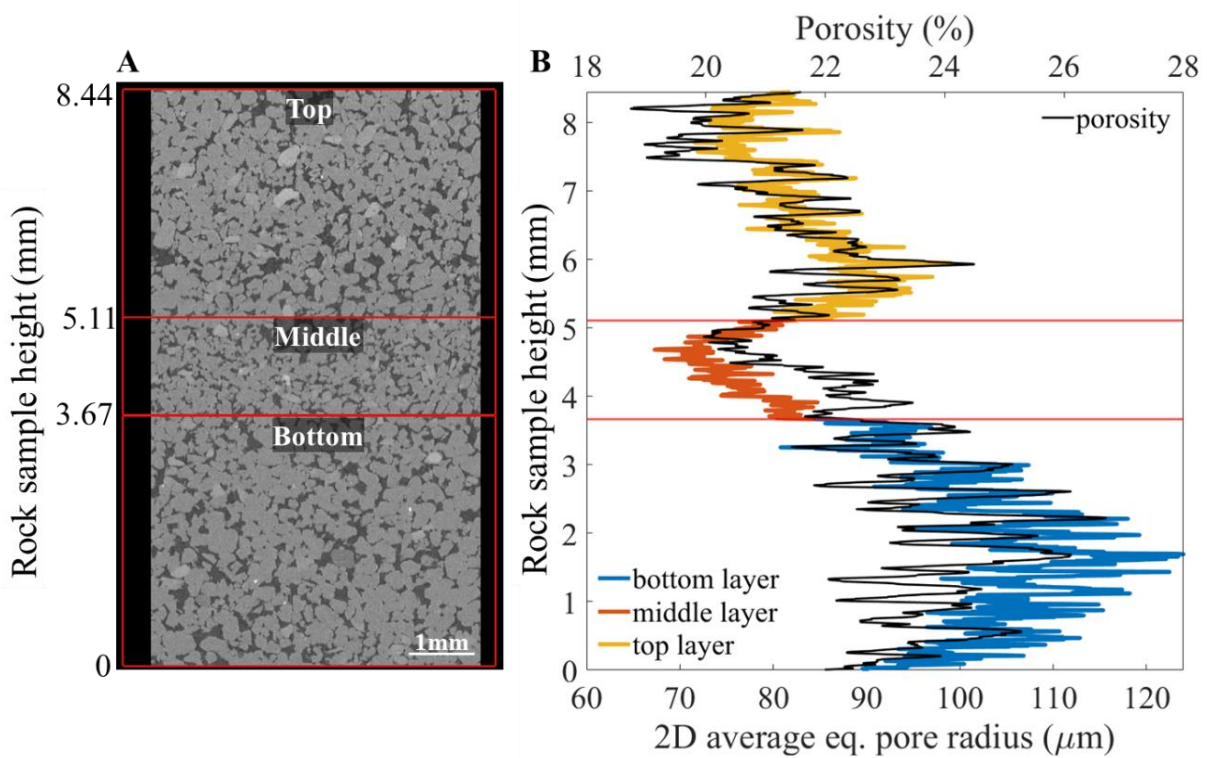

**Figure 2**. Pore and throat size analysis. (A) 2D cross-section showing the rock sample divided into three layers with a visual difference in pore sizes, and (B) average equivalent (eq.) pore radius obtained from 2D average pore area for each 2D vertical slice plotted against the rock sample height. The variation of porosity along the sample is also shown.

## Results and Discussion

*Experiment 1 – Dual flowlines at the bottom of the rock sample*

In this experiment, the bottom of the rock sample was connected to two flowlines, one serving as the outlet during drainage and the other as the inlet during imbibition Figure 1 A&B). Two cycles of drainage and imbibition were performed. During the first drainage step (in Cycle 1), the injected $H_2$ enters the top layer from above, invading the large throats, and stays as a connected phase (Figure 3A).

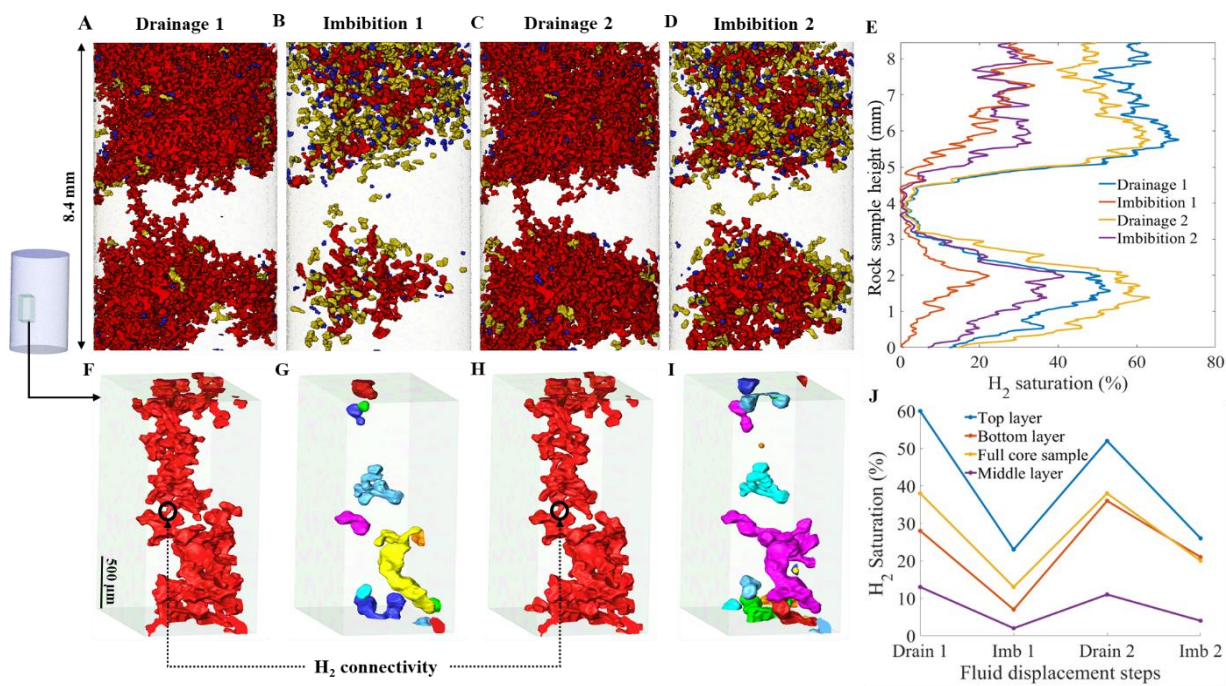

**Figure 3.** $H_2$ saturation analysis (experiment 1). $H_2$ phase visualization inside the full rock sample after each fluid displacement step, (A) after first drainage, (B) after first imbibition, (C) after second drainage, and (D) after second imbibition. In (A) – (D), red colour represents connected $H_2$ phase occupying pore space larger than 100 times the average pore size, yellow colour represents connected $H_2$ phase occupying pore space between 10-100 times the average pore size and blue colour represents connected $H_2$ phase occupying pore space up to 10 times the average pore size, (E) $H_2$ saturation after each fluid displacement step plotted against the rock sample height. $H_2$ phase visualized in a subsection of the rock sample after each fluid displacement step, (F) and (H) after the first and second drainage respectively with the smallest throat providing the $H_2$ connection encircled, (G) and (I) after the first and second imbibition respectively showing trapped, disconnected $H_2$ ganglia after imbibition. (J) $H_2$ saturation values in the different layers of the rock sample after each fluid displacement step.

When $H_2$ reaches the middle layer, the decrease in the number of connected flow paths through larger pores could cause a restriction to flow due to capillary forces,[33] as a higher capillary pressure is needed to invade the middle layer consisting of smaller pores and throats. This flow restriction could result in the accumulation of $H_2$ just above the middle layer, increasing the $H_2$ saturation and the capillary pressure. The $H_2$ saturation after drainage in the full rock sample is 38% compared to a $H_2$ saturation of 60% in the top layer and 28% in the bottom layer. The 3D visualization of $H_2$ after drainage shows that $H_2$ is connected between the top and the bottom layer via a single channel (Figure 3A). This connection indicates that $H_2$ flows through the path which requires the lowest capillary entry pressure within the middle layer. This path consists of throats larger than the average throat size in the middle layer (Figure 4).

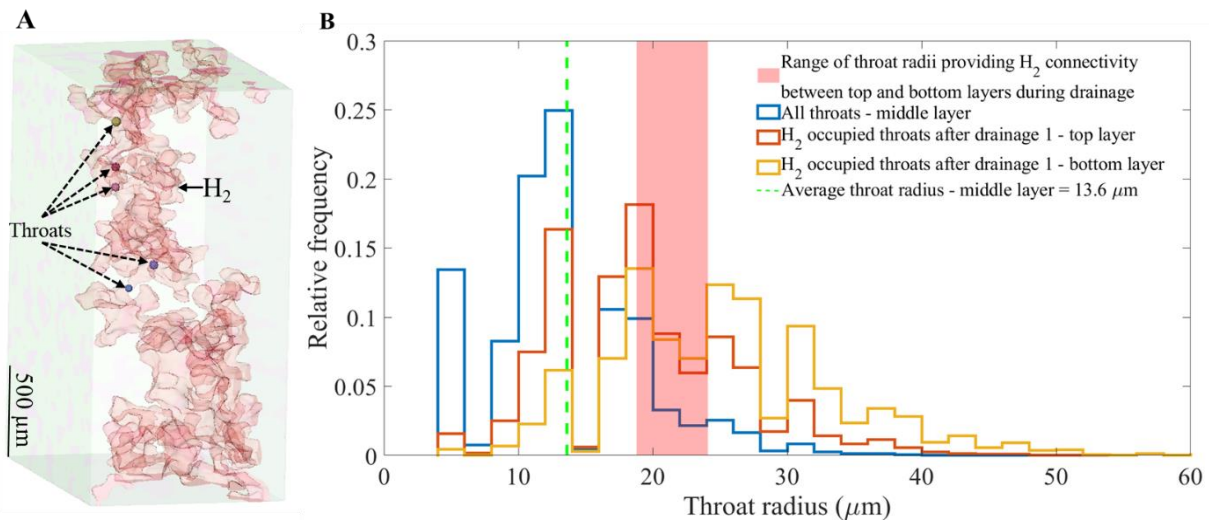

**Figure 4.** Throat size analysis (experiment 1 – drainage). (A) a subsection of the rock sample with throats shown as spheres that allow for connectivity of $H_2$ phase (semi-transparent red) between the top and bottom layers during drainage, (B) throat size distributions of all throats in the middle layer and $H_2$ occupied throats in the top and bottom layers during the first drainage. The average throat radius of all the throats in the middle layer is marked as a dashed-green line and the throats present in the pathway providing the $H_2$ connectivity between the top and bottom layers during drainage are highlighted by the red-shaded region.

The flow of $H_2$ through the channel in the middle layer results in the $H_2$ bypassing most of the middle layer and an increase in $H_2$ saturation just below the middle layer. This observation aligns with findings in a 2D study[34] for an oil-water system, where channeling of the non-wetting phase was reported in a micromodel consisting of layers of varying throat sizes.

Brine is then injected from the bottom of the rock sample to displace the $H_2$. Since there are separate flow lines at the bottom for the $H_2$ outlet during drainage and brine inlet during imbibition, the brine directly enters the rock sample during imbibition and starts displacing $H_2$. The overall residual $H_2$ saturation after imbibition is 13%, showing a recovery factor of 66%. Even though the residual $H_2$ saturation in the top layer is higher compared to the full rock sample (Figure 3J), the recovery trend is comparable. $H_2$ remains as small, disconnected ganglia throughout the rock sample (Figure 3B and Figure 3G) as an increase in the brine phase saturation leads to snap-off and trapping of $H_2$ in the larger pores. $H_2$ phase connectivity between the top and the bottom layer is also broken off with most of the $H_2$ ganglia trapped above and below the middle layer.

$H_2$ is then re-injected into the rock sample from the top (Cycle 2). The connectivity of $H_2$ occurs through the same channel as in the first drainage, bypassing most of the middle layer (Figure 3C, and Figure 3H). The overall $H_2$ saturation of 38% after the second drainage is the same as the first drainage, however, sectional analysis shows that there is a difference in $H_2$ saturations in the top and bottom layers (Figure 3J). $H_2$ saturation in the bottom layer after the second drainage is higher compared to the first drainage, likely due to the presence of residual $H_2$ clusters after the first imbibition.

Brine re-injection from the bottom for the second imbibition follows the second drainage step. Results show that the overall recovery factor of 47% for this cycle is lower compared to the recovery factor for the first cycle (66%). In the top layer, the $H_2$ saturation profiles are similar for both the imbibition steps, with a residual $H_2$ saturation of 23% and 26% after the first imbibition and the second imbibition respectively. However, in the bottom layer, the residual $H_2$ saturation of 21% is considerably higher after the second imbibition compared to 7% after the first imbibition. The higher saturation of $H_2$ in the bottom layer after the second drainage contributes to higher residual trapping of $H_2$ after the second imbibition. 3D visualization of $H_2$ (Figure 3D) shows that the higher residual saturation in the bottom layer is due to the presence of a single large ganglion that contributes to over 70% of the residual $H_2$ saturation in the bottom layer (Figure S6 – Supplementary data). The trapping of a large non-wetting phase ganglion during imbibition, resulting from throat size variation, has also been observed in 2D micromodel experiments for oil-water systems.[34]

Pore and throat occupancy analysis for $H_2$ reveals a preferential filling of larger pores and throats (Figure 5). However, due to the presence of larger pores and throats in the bottom layer (cf. Table 1), the average pore and throat radii of the $H_2$-filled pores in this layer were larger compared to the other layers. The difference in the sizes of the $H_2$ occupied pores and throats in the top layer and the average of the overall pores and throats of the top layer is smaller compared to the bottom layer. This relatively small difference between the average radii of all the pores and throats in the top layer and the $H_2$ occupied pores and throats in the top layer indicates that $H_2$ starts filling relatively smaller pores and throats within the top layer during drainage, which is caused by restriction to $H_2$ flow through the middle layer, until a connection through the middle layer is established for $H_2$ to flow into the bottom layer.

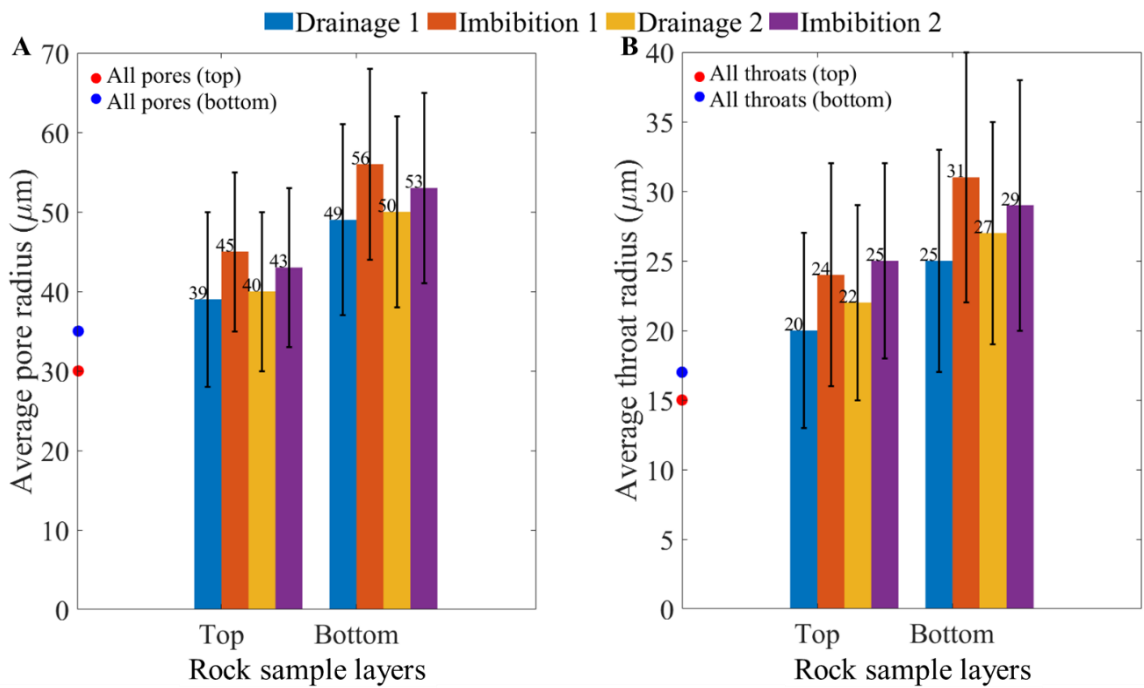

**Figure 5.** (A) Average pore radius of $H_2$ occupied pores and (B) average throat radius of $H_2$ occupied throats after each fluid displacement step in the top and bottom layers of the rock sample. The red and blue filled symbols represent the average pore radius of all the pores in the top and bottom layers in (A) respectively and the average throat radius of all the throats in the top and bottom layers in (B) respectively.

The average radii of the $H_2$ occupied pores and throats increase after imbibition (Figure 5), indicating that brine occupies most of the smaller pores with $H_2$ left in the larger pores. Due to the higher $H_2$ residual saturation in the bottom layer after the second imbibition compared to the first imbibition, $H_2$ enters more of the smaller pores and throats in the bottom layer. As seen in Figure 5B, the increase in the average throat radius of $H_2$ occupied throats for the second imbibition (from 27 to 29 µm) is significantly smaller compared to the first imbibition step (from 25 to 31 µm). Additionally, a much higher number of pores and throats are occupied by $H_2$ after the second imbibition compared to the first imbibition in the bottom layer. The number of pores and throats occupied by $H_2$ after each fluid displacement step in the different layers of the rock sample and the corresponding values of the average pore and throat radii are provided in Table S3 and Table S4 (Supplementary data).

Overall, this cyclic $H_2$ injection and withdrawal experiment shows that the small-scale heterogeneities in the rock have a significant influence on the pore-scale fluid displacement and trapping during drainage and imbibition.

*Experiment 2 – Single flowline at the bottom of the rock sample*

Using the same rock sample, we performed another experiment with a slight modification in the experimental apparatus. In this experiment, a single flowline was connected to the bottom of the rock sample (Figure 1C&D). This flowline served as the outlet from the rock sample during drainage and the inlet to the rock sample during imbition. This type of flowline configuration is mostly used in experiments conducted on homogeneous rock samples in which we observe uniform fluid distribution of the $H_2$ after drainage.[20] When the flow is reversed during imbition, we do not expect the $H_2$ present in the flowline to enter additional pores that have remained filled with brine after drainage. In our experiment on a layered rock sample, due to the variation in the pore and throat sizes in different layers, the fluid configurations vary significantly with the change in injection strategy, and we observe $H_2$ occupying additional pores during imbition (Figure 6).

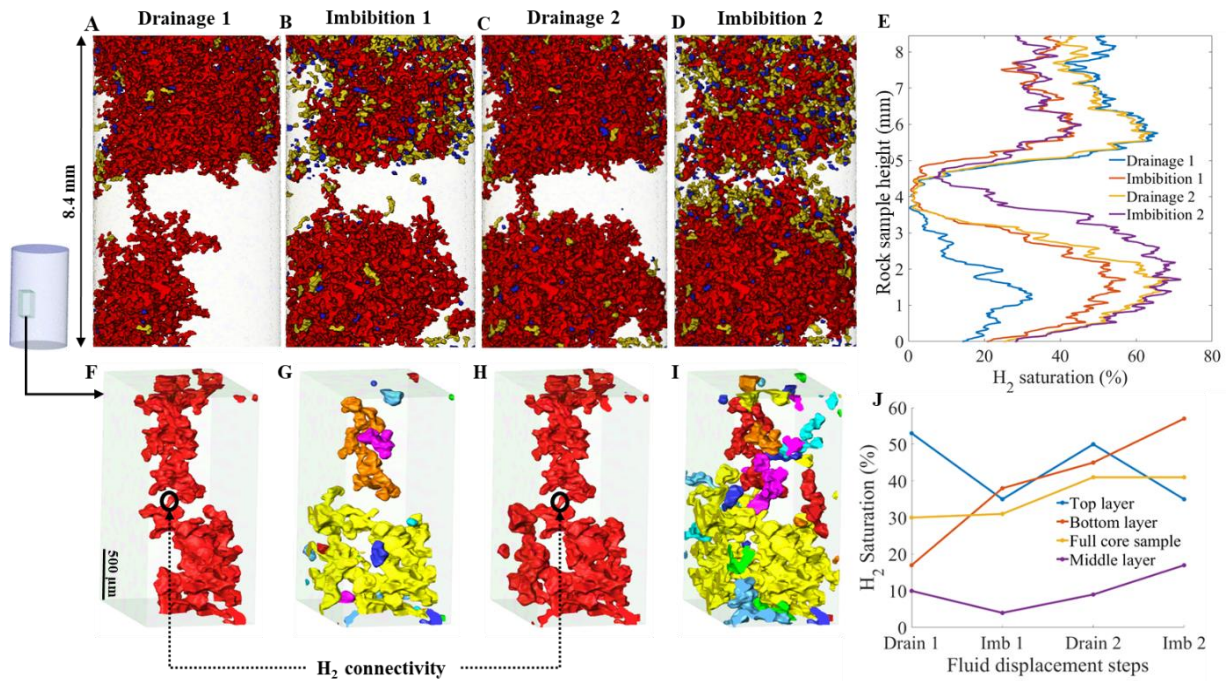

**Figure 6.** $H_2$ saturation analysis (experiment 2). $H_2$ phase visualization inside the full rock sample after each fluid displacement step, (A) after first drainage, (B) after first imbibition, (C) after second drainage, and (D) after second imbibition. In (A) – (D), red colour represents connected $H_2$ phase occupying pore space larger than 100 times the average pore size, yellow colour represents connected $H_2$ phase occupying pore space between 10-100 times the average pore size and blue colour represents connected $H_2$ phase occupying pore space up to 10 times the average pore size, (E) $H_2$ saturation after each fluid displacement step plotted against the rock sample height. $H_2$ phase visualized in a subsection of the rock sample after each fluid displacement step, (F) and (H) after imbibition. (J) $H_2$ saturation values in the different layers of the rock sample after each fluid displacement step.

During the first drainage, $H_2$ fills the top layer of the rock sample (Figure 6A) and bypasses the middle layer after breakthrough. After this drainage, the $H_2$ saturation is 53% in the top layer, while in the bottom layer, the $H_2$ saturation is 17% and confined to one side. The full rock sample has a $H_2$ saturation of 30%, which is lower than that in experiment 1, due to the low saturation in the bottom layer (see Figure 6J). As described for experiment 1 in the previous section, the higher $H_2$ saturation in the top layer is possibly due to the higher local capillary pressure reached during drainage in the section upstream from the middle layer than in the section downstream of the middle layer because the middle layer has a higher capillary invasion pressure.

Subsequently, brine is injected from the bottom of the rock sample during the first imbibition cycle. Following this fluid displacement step, we observe a notable increase in the $H_2$ saturation in the bottom layer of the rock sample, as it rises from 17% (after drainage) to 38% (after imbibition). This increase may be attributed to the presence of $H_2$ within the flowline, introducing an additional volume of $H_2$ into the rock sample from the bottom prior to imbibition. This additional volume of $H_2$ invades more pore space and increases the $H_2$ saturation in the bottom layer as this layer is now upstream to the middle layer, allowing the local capillary pressure in the bottom layer to increase. For a homogeneous sample, this change in the direction of $H_2$ injection would not increase the $H_2$ saturation, as the $H_2$ would flow through the already invaded pore space since there is no mechanism to increase the local capillary pressure.

During imbibition, it is likely that snap-off of the $H_2$ phase occurs in the middle layer (Figure 6G) as it has smaller pores and throats, which are the first places to be invaded by brine. This snap-off results in the disconnection of the $H_2$ phase and forces the $H_2$ to accumulate below the middle layer, creating a drainage-like condition. We observe a slight increase in the overall $H_2$ saturation after the first imbibition which is 31% as compared to the first drainage (30%). The saturation profiles are opposite for the different layers of the rock sample, with $H_2$ saturation reducing from 53% to 35% in the top layer while increasing from 17% to 38% in the bottom layer (see Figure 6E and Figure 6J). As there is no restriction to flow in the top layer, only trapped $H_2$ ganglia are left after imbibition. All the additional $H_2$ accumulates in the bottom layer (Figure 6B), emphasizing the role of the small changes in the pore and throat sizes on the fluid distributions and trapping.

The double displacement, i.e., reverse drainage process during imbibition, observed in the layered sample would not have occurred in a homogeneous rock sample and a similar recovery factor (in a completely homogeneous rock sample) would have been expected as in the top layer. The recovery factor of 34% for the top layer matches closely with the results from our previous experiment conducted on a homogeneous rock sample at the same conditions, using the same experimental strategy.[20]

The second flow cycle is then initiated, and $H_2$ is re-injected from the top. In this drainage step, we observe a similar saturation profile in the top layer as in the first drainage step (Figure 6E). The connecting path between the top and bottom layers remains the same for both drainage cycles (Figure 6F and Figure 6H) with most of the middle layer bypassed by $H_2$. $H_2$ saturation (after the second drainage) in the bottom layer is 45%, which is higher compared to the first drainage due to the high $H_2$ saturation obtained after the first imbibition.

Brine is then re-injected from the bottom for the second imbibition step. We observe that the $H_2$ saturation in the bottom layer further increases after the second imbibition, as additional $H_2$ in the flow line enters the rock sample from the bottom. Similar to that in the first cycle, it is likely that snap-off occurred in the middle layer during imbibition, breaking the $H_2$ connectivity between the top and bottom layers (Figure 6I). The saturation profile in the top layer follows a similar trend to that of the first imbibition (Figure 6E), with only trapped $H_2$ ganglia left in the pore space. For the second imbibition, we observe that $H_2$ invades relatively smaller pores and throats of the bottom layer and even part of the middle layer. The overall $H_2$ saturation is 41% which is the same as the overall saturation after the second drainage. The $H_2$ saturation reduces from 50% to 35% in the top layer and increases from 45% to 57% in the bottom layer after the second imbibition (see Figure 6J). The top layer has the same residual $H_2$ saturation after the second imbibition as after the first imbibition and displays a saturation profile and recovery factor comparable to a homogeneous rock sample.[20]

Overall, it is evident that the pore-scale heterogeneity in our rock sample strongly influences the $H_2$ saturation if the injection strategy is changed. For our rock sample, the top layer could present the closest representation of a homogenous rock, and if analyzed separately, shows that the residual $H_2$ saturation does not increase after multiple cycles as reported in a recent study.[14] However, as observed in this study, small-scale heterogeneities in the pore space of a layered sample have a substantial influence on the residual saturation of $H_2$.

The analysis of the $H_2$ pore occupancy is consistent with the saturation trend in the different layers of the rock sample as shown in Figure 7.

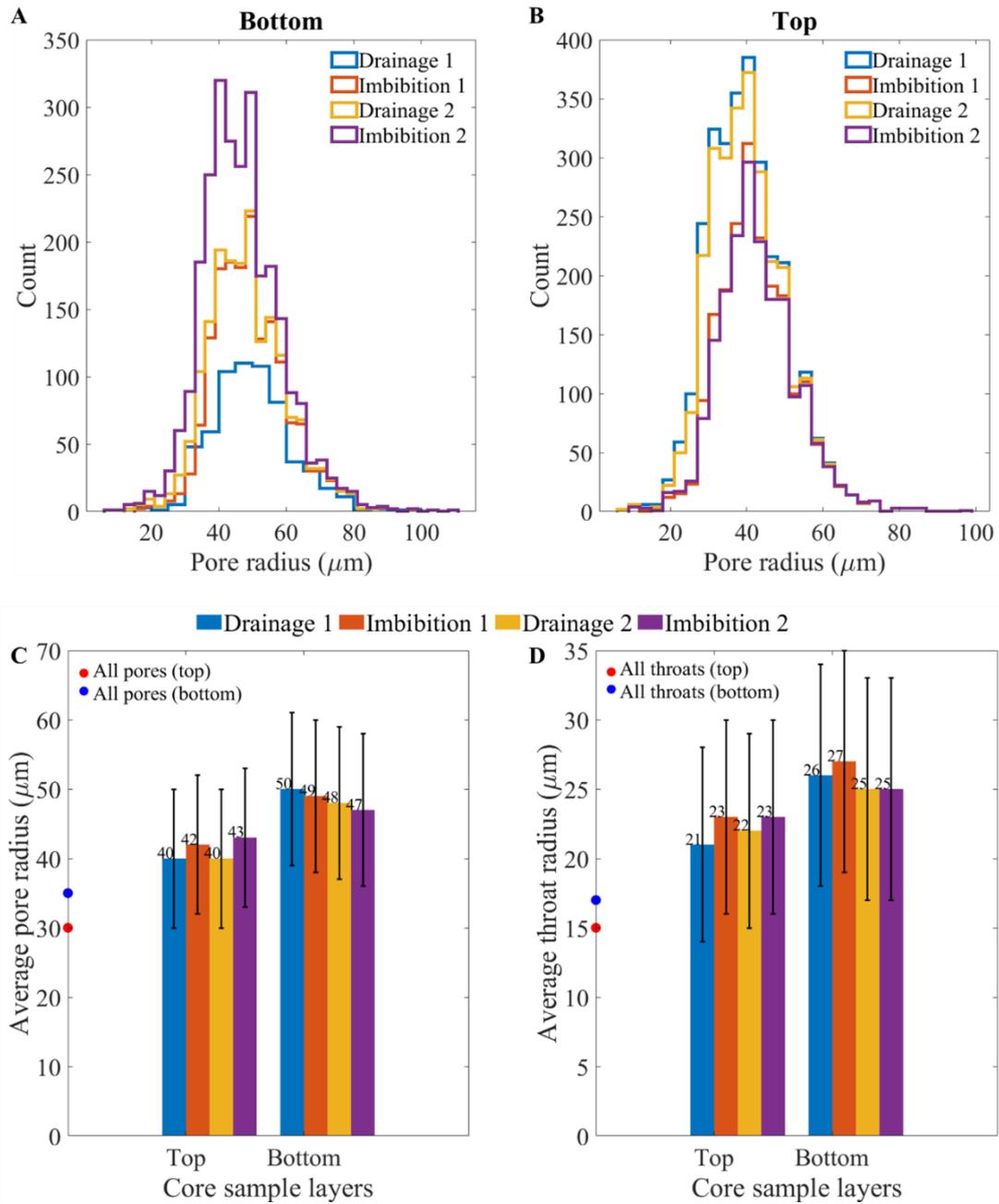

**Figure 7.** Comparison of $H_2$ pore occupancy: (A) bottom layer after each displacement step shows that the number of pores occupied by $H_2$ increases in each subsequent fluid displacement step with the peak shifting towards the left, (B) top layer after each fluid displacement step shows that the number of $H_2$ occupied pores is higher after each drainage compared to each imbibition and the average pore radius of $H_2$ occupied pores is larger for imbibition steps compared to drainage steps. Additionally in the top layer, the plots for both the drainage steps match each other as do the plots for both the imbibition steps, (C) average pore radius of $H_2$ occupied pores and (D) average throat radius of $H_2$ occupied throats after each fluid displacement step in the top and bottom layers of the core sample. The red and blue filled symbols represent the average pore radius of all the pores in the top and bottom layers in (C) respectively and the average throat radius of all the throats in the top and bottom layers in (D) respectively.

In the bottom layer, where $H_2$ saturation consistently increases throughout the fluid displacement steps, there is a corresponding increase in the number of $H_2$ occupied pores and throats from the first drainage up to the second imbibition (Figure 7A). Additionally, as $H_2$ saturation increases with each fluid displacement step in the bottom layer, with $H_2$ invading more smaller pores and throats, we observe a decreasing trend in the average pore and throat radii of $H_2$ occupied pores and throats after each fluid displacement step (Figure 7C and Figure 7D). The number of pores and throats occupied by $H_2$ after each fluid displacement step in the different layers of the rock sample and the corresponding values of the average pore and throat radii are provided in Table S5 and Table S6 (Supplementary data).

Conversely, in the top layer, we observe that the number of $H_2$ occupied pores and throats is less after each imbibition step compared to the previous drainage step, and the average pore and throat radii of $H_2$-occupied pores and throats is higher after each imbibition step compared to the previous drainage step (Figure 7B, Figure 7C and Figure 7D), which is similar to that observed in a homogeneous sample.[20]

Pressure drop data measured across the rock sample shows spikes during drainage (Figure S7A – Supplementary data) which could be due to the flow restrictions in the middle layer of the rock sample. However, due to the small pore volume of the rock sample, the exact time when $H_2$ reached the middle layer is difficult to ascertain. A higher pressure drop observed during the second imbibition (Figure S7B – Supplementary data) is indicative of $H_2$ filling the middle layer, as we observe a higher saturation of $H_2$ in the middle layer after the second imbibition (Figure 6D, Figure 6E, and Figure 6J).

## Conclusions

Through this study, we offer valuable insights into the impact of pore-scale heterogeneity on $H_2$ displacement and trapping in subsurface reservoirs. We show that for low flowrate (capillary dominated) displacement experiments, capillary pressure effects complicate the fluid displacements, particularly near heterogeneous boundaries, and significantly influence the initial $H_2$ saturation and the subsequent trapping of $H_2$. We found that $H_2$ flows along a preferential pathway through a low permeable layer, bypassing a section of the rock sample, and thereby reducing the initial storage capacity. Our results also show that during $H_2$ withdrawal from a heterogeneous rock, a larger volume of $H_2$ could get trapped below low permeability layers due to snap-off events occurring in the smaller throats of the low permeability layer.

These findings show that heterogenous rocks behave differently from homogenous rocks. The small heterogeneities in pore and throat sizes, and permeabilities analyzed in this study show a surprising effect on flow, saturation, and trapping of $H_2$ in reservoir rocks. The impact of small-scale heterogeneities on UHS should be studied further using different capillary numbers and permeability contrasts. Furthermore, we emphasize the importance of the direction of fluid injection on $H_2$ distribution and trapping in a heterogeneous rock, whereby if any volume of $H_2$ enters the rock sample before the imbibing brine, the residual $H_2$ saturation could increase after imbibition compared to the initial $H_2$ saturation after drainage.

Time-resolved 3D visualization experiments could provide further insights into the pore-scale dynamics occurring within the heterogeneous layers of the rock sample. Understanding such pore-scale phenomena is crucial for informing and validating pore-scale models. While this work focuses on the pore-scale, it highlights the importance of considering small-scale heterogeneity, which is often overlooked in large-scale reservoir models, when designing and implementing UHS systems.

## Data Availability Statement

The raw, filtered, and segmented tomographic datasets and the extracted pore and throat image files have been uploaded to a public repository (zenodo), the details of which are provided at https://doi.org/10.5281/zenodo.8375683.

## Acknowledgement

We thank Helen Lewis for providing access to the X-ray tomography facility at the Institute of GeoEnergy Engineering at Heriot-Watt University and Mehran Sohrabi for providing the rock for these experiments. We gratefully acknowledge Jim Buckman, Clayton Magill, Paul Miller, Robert Louden, and Jim Allison for their support in the preparation of the equipment and materials used for our experiment. We are also thankful to Energi Simulation for providing partial funding for this research.

**Supplementary data**

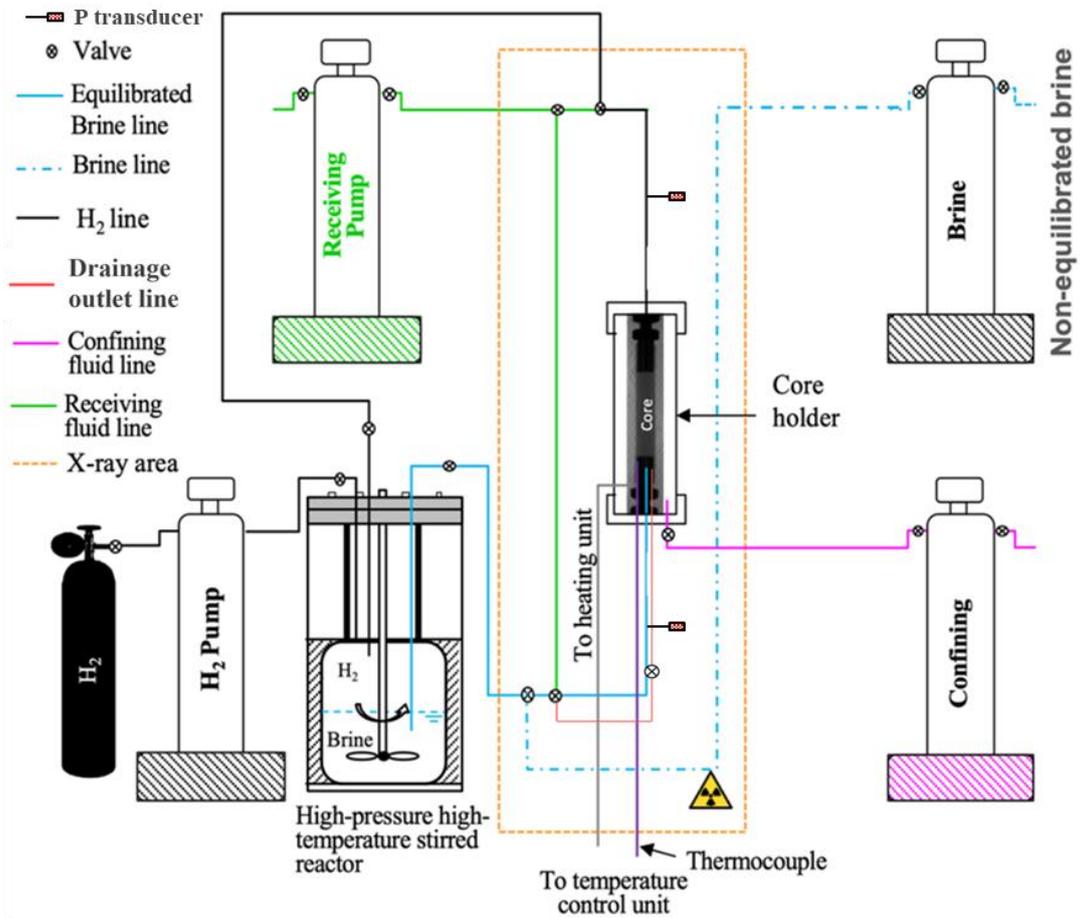

**Figure S1.** Schematic of the flow system. An additional outlet line was added at the bottom of the rock sample for experiment 1, marked as the 'Drainage outlet line'.

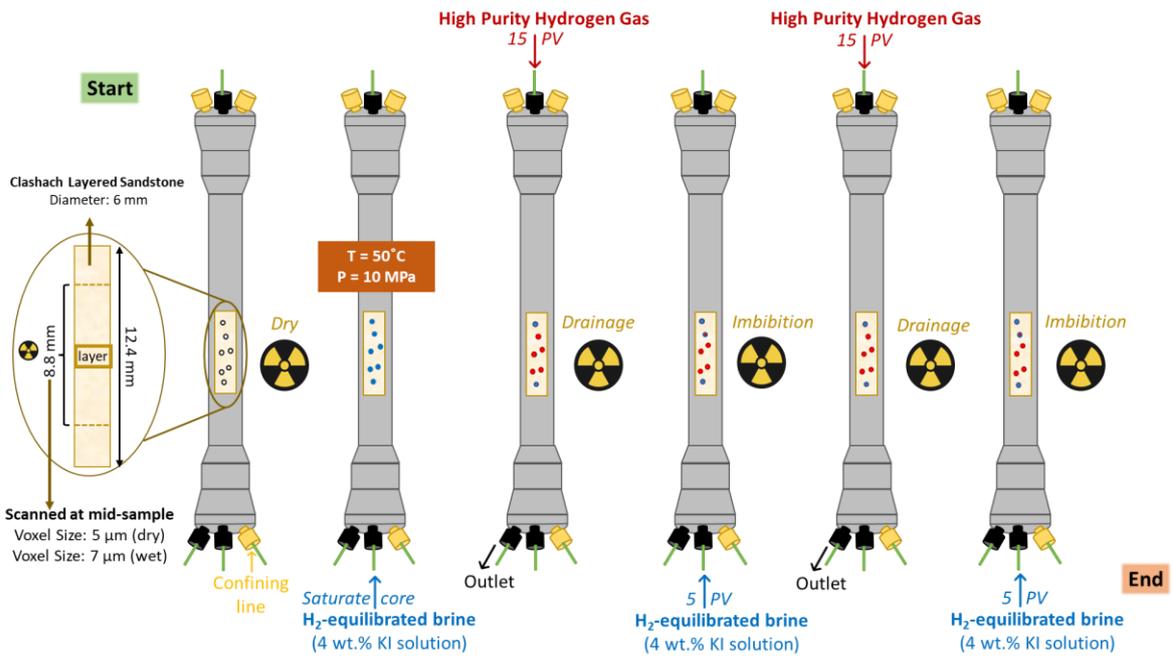

**Figure S2.** Experimental steps (experiment 1); the rock sample was initially scanned as dry and then after each fluid displacement step.

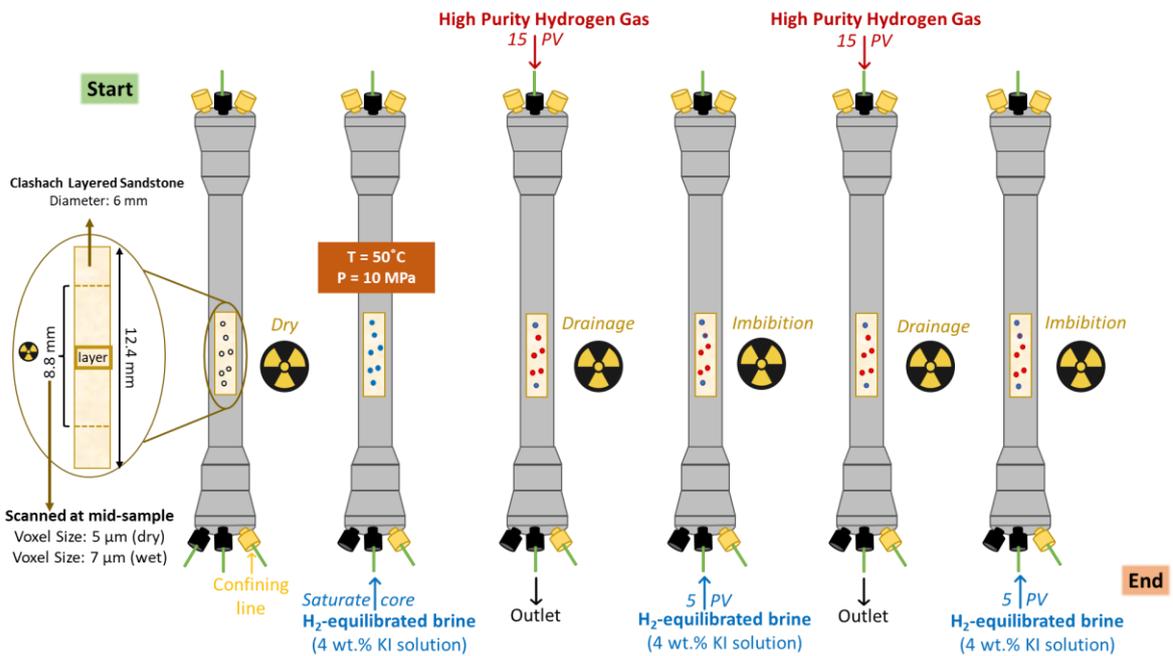

**Figure S3.** Experimental steps (experiment 2); the rock sample was initially scanned as dry and then after each fluid displacement step.

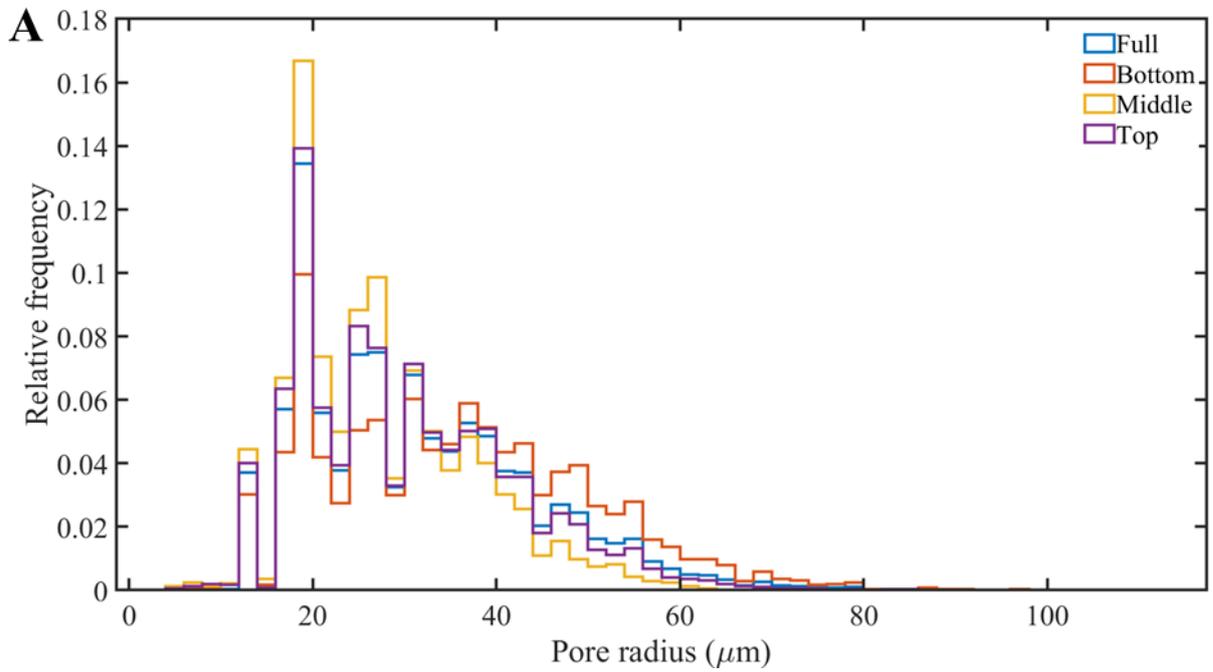

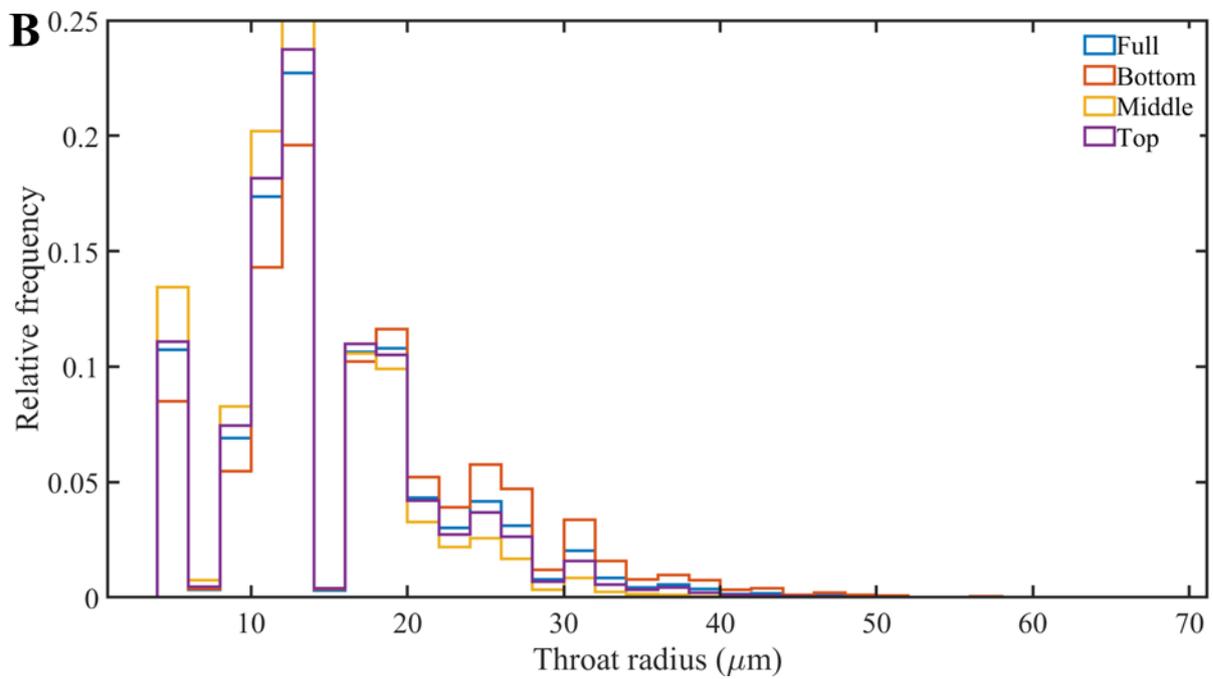

**Figure S4.** (A) pore size distribution and (B) throat size distribution of all the pores and throats in the different layers of the rock sample

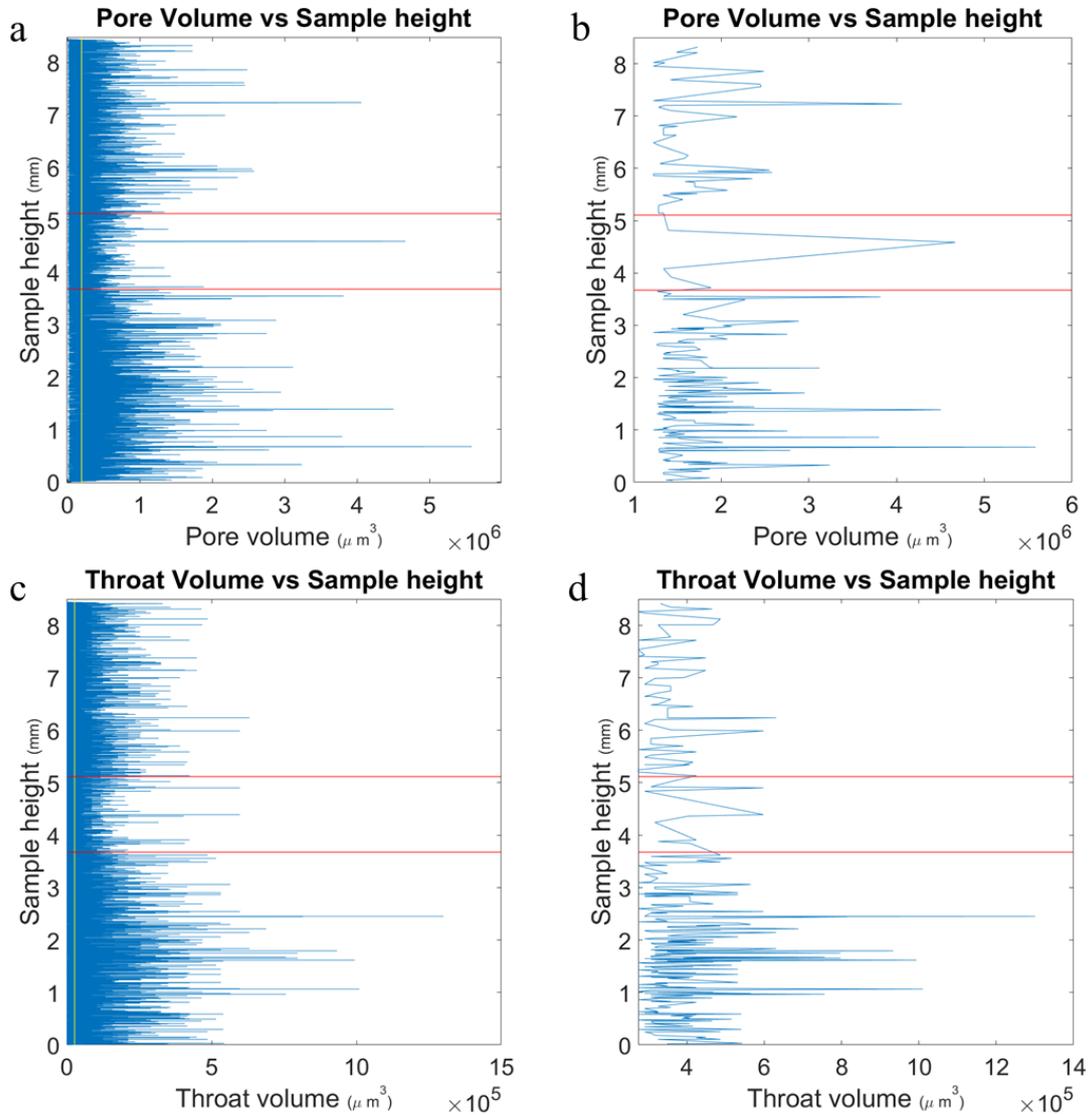

**Figure S5.** 3D pore and throat size analysis. (A) all pore volumes plotted against the rock sample height, (B) pores with a volume more than 6 times the average pore volume plotted against the rock sample height (to highlight the presence of only the large pores throughout the rock sample), (C) all throat volumes plotted against the sample height, (D) throats with a volume more than 10 times the average throat volume plotted against the rock sample height (to highlight the presence of only the large throats throughout the rock sample). The red horizontal lines in all images identify the middle layer, while the yellow vertical line in (A) and (C) mark the average pore and throat volume respectively.

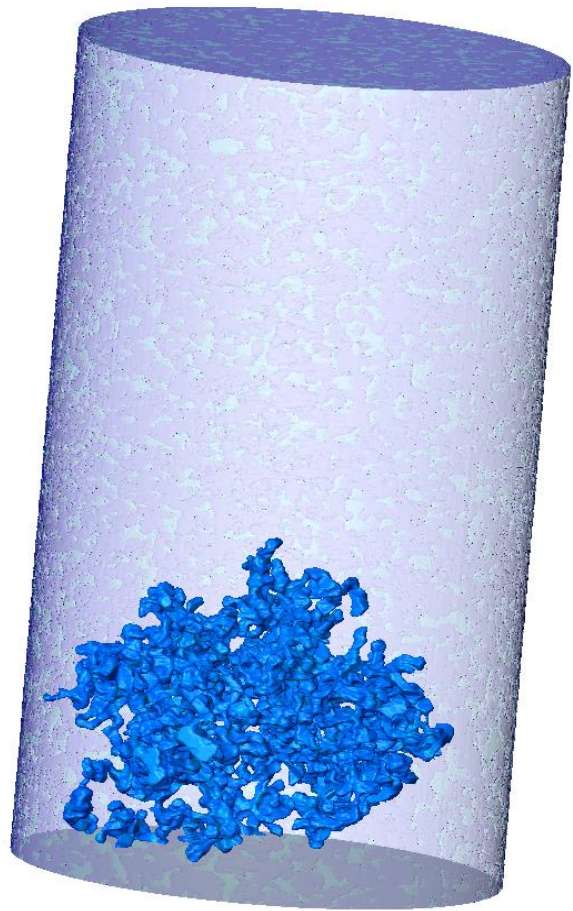

**Figure S6.** Experiment 1 (imbibition 2); the large connected $H_2$ ganglia trapped in the bottom layer after the second imbibition

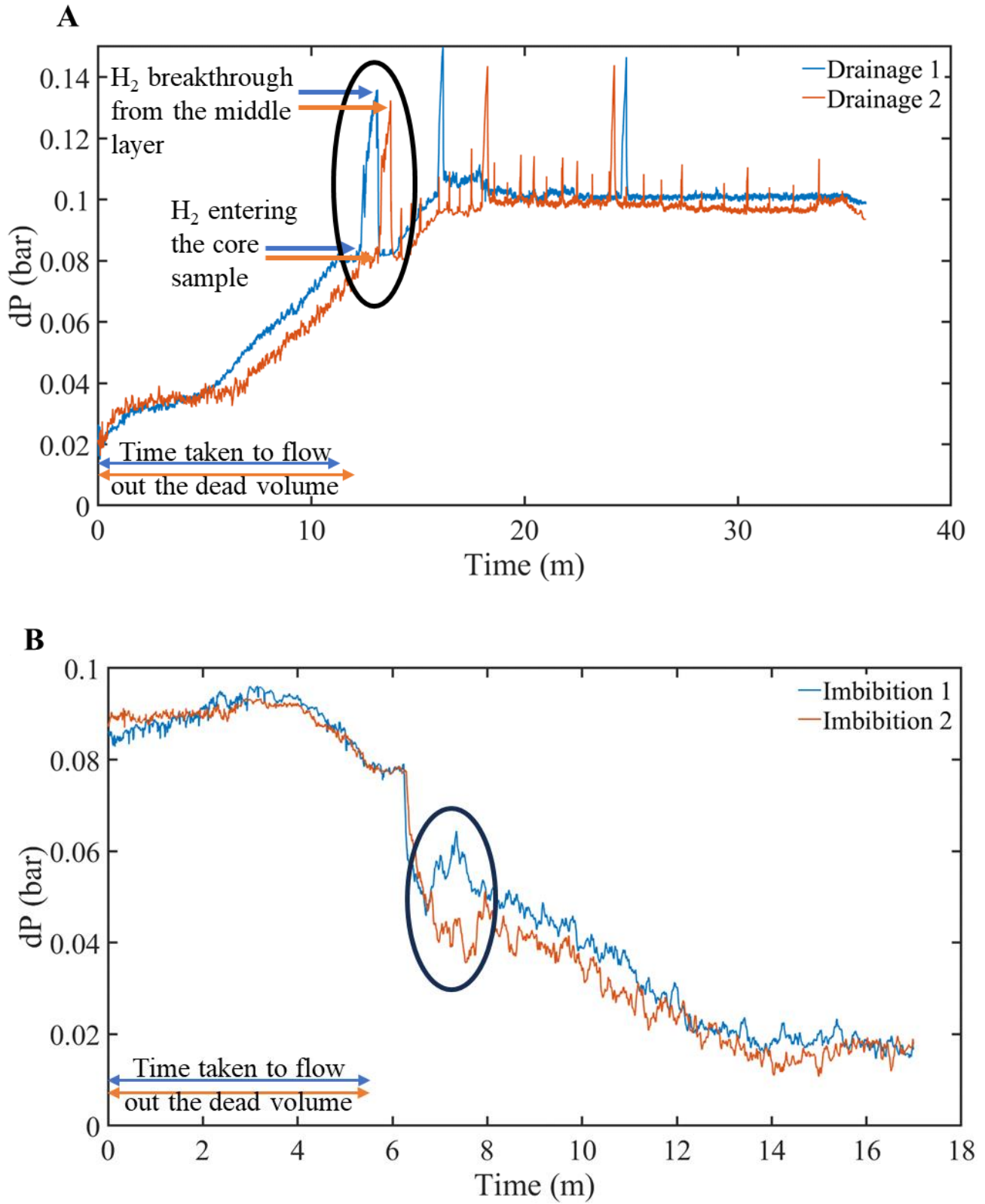

**Figure S7.** Pressure drop profiles (experiment 2). (A) pressure drop during the drainage steps with the estimated time when $H_2$ enters the rock sample and breakthroughs from the middle layer, (B) pressure drop during the imbibition steps with the difference in pressure drop between the two imbibition steps encircled

**Table S1.** X-ray scanning parameters for all the scans acquired for the two experiments.

| X-ray equipment | RX Solutions EasyTom 150 X-ray micro-CT machine |
|---|---|
| X-ray tube power | 10 W |
| X-ray tube voltage | 120 kV |
| X-ray tube current | 83 uA |
| No of projections per scan | 1792 |
| Voxel size | 5 µm (dry scan), 7 µm (all wet scans) |

**Table S2.** Grayscale intensity threshold values used for the segmentation of the pores in the image from the dry scan and $H_2$ from the images of the wet scans using the watershed algorithm in Avizo 2022 software. The gradient values were automatically selected by the algorithm.

|  |  |  | **Grayscale intensity threshold range** |
|---|---|---|---|
|  | Dry scan | Pores | 12000 – 18665 |
| **Experiment 1** | Drainage 1 | $H_2$ | 11111 – 17700 |
|  | Imbibition 1 |  | 11815 – 17950 |
|  | Drainage 2 |  | 10488 – 17470 |
|  | Imbibition 2 |  | 10887 – 17500 |
| **Experiment 2** | Drainage 1 |  | 14000 – 17100 |
|  | Imbibition 1 |  | 14000 – 16850 |
|  | Drainage 2 |  | 14000 – 16800 |
|  | Imbibition 2 |  | 14000 – 16615 |

The images were filtered prior to segmentation using a non-local means filter with the following settings:

| Images | Mode | Search Window [px] | Local neighborhood [px] | Similarity Value |
|---|---|---|---|---|
| Dry scan and Experiment 1 | GPU Standard | 5 | 3 | 0.1 |
| Experiment 2 | GPU Standard | 4 | 3 | 0.1 |

**Table S3.** Count of the H$_2$ occupied pores and throats after each fluid displacement step in all layers of the rock sample (experiment 1)

|   |   | 5.11-8.44 mm | 3.67-5.11 mm | 0-3.67 mm | 0-8.44 mm |
|---|---|---|---|---|---|
|   |   | 477 slices | 206 slices | 524 slices | 1207 slices |
|   |   | Top | Middle | Bottom | Full |
|   |   | # of pores / throats | | | |
| **Drainage 1** | **Pores** | 3198 | 319 | 1128 | 4577 |
|   | **Throats** | 7714 | 723 | 2835 | 11206 |
| **Imbibition 1** | **Pores** | 1349 | 58 | 338 | 1729 |
|   | **Throats** | 1665 | 61 | 498 | 2218 |
| **Drainage 2** | **Pores** | 2930 | 295 | 1406 | 4570 |
|   | **Throats** | 5556 | 508 | 3100 | 9111 |
| **Imbibition 2** | **Pores** | 1600 | 116 | 862 | 2554 |
|   | **Throats** | 2023 | 135 | 1461 | 3602 |

**Table S4.** Average pore and throat radii for the H$_2$ occupied pores and throats after each fluid displacement step in all layers of the rock sample (experiment 1)

|   |   | 5.11-8.44 mm | 3.67-5.11 mm | 0-3.67 mm | 0-8.44 mm |
|---|---|---|---|---|---|
|   |   | 477 slices | 206 slices | 524 slices | 1207 slices |
|   |   | Top | Middle | Bottom | Full |
|   |   | µm | | | |
| **Drainage 1** | **Avg. pore radius** | 39 | 37 | 49 | 42 |
|   | **Avg. throat radius** | 20 | 19 | 25 | 21 |
| **Imbibition 1** | **Avg. pore radius** | 45 | 43 | 56 | 47 |
|   | **Avg. throat radius** | 24 | 25 | 31 | 26 |
| **Drainage 2** | **Avg. pore radius** | 40 | 38 | 50 | 43 |
|   | **Avg. throat radius** | 22 | 21 | 27 | 23 |
| **Imbibition 2** | **Avg. pore radius** | 43 | 42 | 53 | 47 |
|   | **Avg. throat radius** | 25 | 24 | 29 | 27 |

**Table S5.** Count of the H₂ occupied pores and throats after each fluid displacement step in all layers of the rock sample (experiment 2)

|  |  | 5.11-8.44 mm<br>477 slices<br>Top | 3.67-5.11 mm<br>206 slices<br>Middle | 0-3.67 mm<br>524 slices<br>Bottom | 0-8.44 mm<br>1207 slices<br>Full |
|---|---|---|---|---|---|
|  |  | # of pores / throats | | | |
| **Drainage 1** | **Pores** | 2938 | 258 | 619 | 3757 |
|  | **Throats** | 5851 | 464 | 1574 | 7838 |
| **Imbibition 1** | **Pores** | 2036 | 131 | 1651 | 3784 |
|  | **Throats** | 3020 | 164 | 3168 | 6338 |
| **Drainage 2** | **Pores** | 2800 | 251 | 1797 | 4789 |
|  | **Throats** | 5224 | 415 | 4333 | 9930 |
| **Imbibition 2** | **Pores** | 1960 | 558 | 2633 | 5071 |
|  | **Throats** | 3067 | 813 | 5434 | 9267 |

**Table S6.** Average pore and throat radii for the H₂ occupied pores and throats after each fluid displacement step in all layers of the rock sample (experiment 2)

|  |  | 5.11-8.44 mm<br>477 slices<br>Top | 3.67-5.11 mm<br>206 slices<br>Middle | 0-3.67 mm<br>524 slices<br>Bottom | 0-8.44 mm<br>1207 slices<br>Full |
|---|---|---|---|---|---|
|  |  | µm | | | |
| **Drainage 1** | **Avg. pore radius** | 40 | 38 | 50 | 42 |
|  | **Avg. throat radius** | 21 | 21 | 26 | 22 |
| **Imbibition 1** | **Avg. pore radius** | 42 | 40 | 49 | 46 |
|  | **Avg. throat radius** | 23 | 23 | 27 | 25 |
| **Drainage 2** | **Avg. pore radius** | 40 | 37 | 48 | 43 |
|  | **Avg. throat radius** | 22 | 21 | 25 | 23 |
| **Imbibition 2** | **Avg. pore radius** | 43 | 39 | 47 | 45 |
|  | **Avg. throat radius** | 23 | 22 | 25 | 24 |